\documentclass[graybox,envcountchap,sectrefs,natbib]{svmono}
\usepackage{mathptmx} % defines TimesRoman as standard font, enhanced by symbols
\usepackage{helvet}       % defines Helvetica as sans-serif standard font
\usepackage{courier}    % defines Courier as typewriter font (verbatim environments)
\usepackage{type1cm}  % enhances the substitution of TimesRoman on systems which may not have it installed
\usepackage{amsfonts}% allows to use AMS fonts ; added by RoD
\usepackage{amssymb}% allows to use AMS fonts ; added by RoD
\usepackage{xspace}
\usepackage{chapterbib}% allows to use bibliography per chapter
\usepackage{makeidx}         % allows index generation
\usepackage{multicol}        % used for the two-column index, handles proper closure of multi-column sections in general.
\usepackage[bottom]{footmisc}% places footnotes at page bottom
\usepackage{graphicx}        % standard LaTeX graphics tool  % when including figure files of non-pdf/png/jpg type (such as eps)

\begin{document}
%\author{  }
%\title{Astronomy with Radioactivities}
%\subtitle{An Introduction to Astrophysics with Decaying Isotopes}
%\maketitle
%\frontmatter%%%%%%%%%%%%%%%%%%%%%%%%%%%%%%%%%%%%%%%%%%%%%%%%%%%%%%
%\include{dedic}
%\include{foreword}
%\include{preface}
%\include{acknow}
%\tableofcontents
\mainmatter%%%%%%%%%%%%%%%%%%%%%%%%%%%%%%%%%%%%%%%%%%%%%%%%%%%%%%%
\setcounter{page}{0} % Ch1-20 Ch2-76 Ch3- 150 Ch4-    Ch5- Ch7: 286
\setcounter{chapter}{0} %

%%%%%%%%%%%%%%%%%%%%%%%%%%%%%%%%%%%%%%%%%%%%%%%%%
% Macro collection for AwR Book
% 26Nov09: RoD: collect from various chapter authors as used by them
% 07Dec09: RoD: eliminate redundant macros and standardize; avoid font changes
%%%%%%%%%%%%%%%%%%%%%%%%%%%%%%%%%%%%%%%%%%%%%%%%%
% Contents:
% - Journal abbreviations
% - Isotopes
% - astronomical standard constants and units
%
%%%%%%%%%%%%%%%%%%%%%%%%%%%%%%%%%%%%%%%%%%%%%%%%%%%
%
%	\let\jnl@style=\rmfamily
%	\def\ref@jnl#1{{\jnl@style#1}}%
\let\jnl=\rmfamily
\def\refe@jnl#1{{\jnl#1}}%

\newcommand\aj{\refe@jnl{AJ}}%
          % Astronomical Journal
\newcommand\actaa{\refe@jnl{Acta Astron.}}%
  % Acta Astronomica
\newcommand\araa{\refe@jnl{ARA\&A}}%
          % Annual Review of Astron and Astrophys
\newcommand\apj{\refe@jnl{ApJ}}%
          % Astrophysical Journal
\newcommand\apjl{\refe@jnl{ApJ}}%
          % Astrophysical Journal, Letters
\newcommand\apjs{\refe@jnl{ApJS}}%
          % Astrophysical Journal, Supplement
\newcommand\ao{\refe@jnl{Appl.~Opt.}}%
          % Applied Optics
\newcommand\apss{\refe@jnl{Ap\&SS}}%
          % Astrophysics and Space Science
\newcommand\aap{\refe@jnl{A\&A}}%
          % Astronomy and Astrophysics
\newcommand\aapr{\refe@jnl{A\&A~Rev.}}%
          % Astronomy and Astrophysics Reviews
\newcommand\aaps{\refe@jnl{A\&AS}}%
          % Astronomy and Astrophysics, Supplement
\newcommand\azh{\refe@jnl{AZh}}%
          % Astronomicheskii Zhurnal
\newcommand\gca{\refe@jnl{GeoCh.Act}}%
          % Geochimica Acta
\newcommand\grl{\refe@jnl{Geo.Res.Lett.}}%
          % Geoph Res Lett
\newcommand\jgr{\refe@jnl{J.Geoph.Res.}}%
          % Journal of Geophysical Research
\newcommand\memras{\refe@jnl{MmRAS}}%
          % Jrnl RAS Canada
\newcommand\jrasc{\refe@jnl{J.RoySocCan}}%
          % Memoirs of the RAS
\newcommand\mnras{\refe@jnl{MNRAS}}%
          % Monthly Notices of the RAS
\newcommand\na{\refe@jnl{New A}}%
  % New Astronomy
\newcommand\nar{\refe@jnl{New A Rev.}}%
  % New Astronomy Review
\newcommand\pra{\refe@jnl{Phys.~Rev.~A}}%
          % Physical Review A: General Physics
\newcommand\prb{\refe@jnl{Phys.~Rev.~B}}%
          % Physical Review B: Solid State
\newcommand\prc{\refe@jnl{Phys.~Rev.~C}}%
          % Physical Review C
\newcommand\prd{\refe@jnl{Phys.~Rev.~D}}%
          % Physical Review D
\newcommand\pre{\refe@jnl{Phys.~Rev.~E}}%
          % Physical Review E
\newcommand\prl{\refe@jnl{Phys.~Rev.~Lett.}}%
          % Physical Review Letters
\newcommand\pasa{\refe@jnl{PASA}}%
  % Publications of the Astron. Soc. of Australia
\newcommand\pasp{\refe@jnl{PASP}}%
          % Publications of the ASP
\newcommand\pasj{\refe@jnl{PASJ}}%
          % Publications of the ASJ
\newcommand\skytel{\refe@jnl{S\&T}}%
          % Sky and Telescope
\newcommand\solphys{\refe@jnl{Sol.~Phys.}}%
          % Solar Physics
\newcommand\sovast{\refe@jnl{Soviet~Ast.}}%
          % Soviet Astronomy
\newcommand\ssr{\refe@jnl{Space~Sci.~Rev.}}%
          % Space Science Reviews
\newcommand\nat{\refe@jnl{Nature}}%
          % Nature
\newcommand\iaucirc{\refe@jnl{IAU~Circ.}}%
          % IAU Cirulars
\newcommand\aplett{\refe@jnl{Astrophys.~Lett.}}%
          % Astrophysics Letters and Communications
\newcommand\apspr{\refe@jnl{Astrophys.~Space~Phys.~Res.}}%
          % Astrophysics Space Physics Research
\newcommand\nphysa{\refe@jnl{Nucl.~Phys.~A}}%
          % Nuclear Physics A
\newcommand\physrep{\refe@jnl{Phys.~Rep.}}%
          % Physics Reports
\newcommand\procspie{\refe@jnl{Proc.~SPIE}}%
          % Proceedings of the SPIE

 %%%%% Personal Macros %%%%%%%%%%%%%%%%%%%
 %     Isotopes     
\newcommand{\Al}{$^{26}$Al\xspace}
\newcommand{\al}{$^{26}$Al\xspace}
\newcommand{\Be}{$^{7}$Be\xspace}
\newcommand{\be}{$^{7}$Be\xspace}
\newcommand{\bem}{$^{10}$Be\xspace}
\newcommand{\ca}{$^{44}$Ca\xspace}
\newcommand{\Ca}{$^{44}$Ca\xspace}
\newcommand{\cam}{$^{41}$Ca\xspace}
\newcommand{\Co}{$^{56}$Co\xspace}
\newcommand{\co}{$^{56}$Co\xspace}
\newcommand{\csm}{$^{135}$Cs\xspace}
\newcommand{\ct}{$^{13}$C\xspace}
\newcommand{\ci}{$^{57}$Co\xspace}
\newcommand{\Ci}{$^{57}$Co\xspace}
\newcommand{\ch}{$^{60}$Co\xspace}
\newcommand{\Ch}{$^{60}$Co\xspace}
\newcommand{\Cl}{$^{36}$Cl\xspace}
\newcommand{\li}{$^{7}$Li\xspace}
\newcommand{\Li}{$^{7}$Li\xspace}
\newcommand{\Fe}{$^{60}$Fe\xspace}
\newcommand{\fh}{$^{60}$Fe\xspace}
\newcommand{\fe}{$^{56}$Fe\xspace}
\newcommand{\Fr}{$^{57}$Fe\xspace}
\newcommand{\fr}{$^{57}$Fe\xspace}
\newcommand{\mg}{$^{26}$Mg\xspace}
\newcommand{\Mg}{$^{26}$Mg\xspace}
\newcommand{\mn}{$^{54}$Mn\xspace}
\newcommand{\Na}{$^{22}$Na\xspace}
\newcommand{\Ne}{$^{22}$Ne\xspace}
\newcommand{\Ni}{$^{56}$Ni\xspace}
\newcommand{\nh}{$^{60}$Ni\xspace}
\newcommand{\Nh}{$^{60}$Ni\xspace}
\newcommand\nuk[2]{$\rm ^{\rm #2} #1$}  % from Maria Lugaro, Ch3
\newcommand{\pd}{$^{107}$Pd\xspace}
\newcommand{\pb}{$^{205}$Pb\xspace}
\newcommand{\tc}{$^{99}$Tc\xspace}
\newcommand{\Sc}{$^{44}$Sc\xspace}
\newcommand{\Ti}{$^{44}$Ti\xspace}
\newcommand{\ti}{$^{44}$Ti\xspace}
%%%%%%%%%%%%%%%%%%%%%%%%%%%%%%%%%%%%%%%%%
% generic abbreviations 
\def\aa{$\alpha$}
\newcommand{\about}{$\simeq$}
\newcommand{\cms}{cm\ensuremath{^{-2}} s\ensuremath{^{-1}}\xspace}
\newcommand{\degree}{$^{\circ}$}
\newcommand{\flux}{ph~cm\ensuremath{^{-2}} s\ensuremath{^{-1}}\xspace}
\newcommand{\fluxrad}{ph~cm$^{-2}$s$^{-1}$rad$^{-1}$\ }
\newcommand{\ga}{\ensuremath{\gamma}}
\newcommand{\gam}{\ensuremath{\gamma}}
\def\nn{$\nu$}
\def\ra{$\rightarrow$}
%%%%%%%%%%%%%%%%%%%%%%%%%%%%%%%%%%%%%%%%%
%  astronomical constants
\newcommand{\Msol}{M\ensuremath{_\odot}\xspace}
\newcommand{\msol}{M\ensuremath{_\odot}\xspace}
\newcommand{\Msolppc}{M\ensuremath{_\odot} pc\ensuremath{^{-2}}{\xspace}}
\newcommand{\Msolpy}{M\ensuremath{_\odot} y\ensuremath{^{-1}}{\xspace}}
\newcommand{\msb}{M\ensuremath{_\odot}\xspace}
\newcommand{\Msun}{M\ensuremath{_\odot}\xspace}
\newcommand{\Rsun}{R\ensuremath{_\odot}\xspace}
\newcommand{\rsun}{R\ensuremath{_\odot}\xspace}
\newcommand{\Lsun}{L\ensuremath{_\odot}\xspace}
\newcommand{\lsun}{L\ensuremath{_\odot}\xspace}
\newcommand{\solar}{\ensuremath{_\odot}\xspace}
\newcommand{\zs}{Z\ensuremath{_\odot}\xspace}

%%%% conflicting but used so far somewhere?
%\def\cs{cm$^{-2}$ s$^{-1}$}
%%%%%%%%%%%%%%%%

% end personal macros and abbreviations

 % Macro Collection

%%%%%
%%%%%
%%%%%%% edit here: UN-COMMENT your own Chapter here !!!
%%%%%%% edit here: UN-COMMENT your own Chapter here !!!
%%%%%%% edit here: UN-COMMENT your own Chapter here !!!
%%%%%%% edit here: UN-COMMENT your own Chapter here !!!

%\include{part_1} % Intro
  %%%%%%%%%%%%%%%%%%%%% chapter_1.tex %%%%%%%%%%%%%%%%%%%%%%%%%%%%%%%%%
%
% Astronomy with Radioactivities: Chapter 1
%
% 01Apr2008 R.Diehl: Draft
% 22Feb2009 R.Diehl: update, more detail
% 29Mayb2009 R.Diehl: more trimming of text & style
% 26Nov2009: RDiehl: change figure locations to ch1/...
% 14Jan2010: RDiehl: Implement Dec review comments: 
% 24Jan2010: D. Hartmann: Language Editing
% 30Jan2010: RDiehl: Insert DDC's a/b/gam processes; insert abundance definition
% 02Feb2010: RDiehl: correct Boltzmann distribution and its discussion
% 05Feb2010 RDiehl: correct mass fraction def
% 01Mar2010 RDiehl: correct a DH line-break typo, a few words in last paragraph, and add element 112 official naming
% 03Mar2010 RDiehl: add spectrum figure and its text; add subject index
% 04Mar2010 RDiehl: some typo corrections
% 15Mar2010: add afiliation
% 24Mar2010: change p to n decay equation, eliminate 'phase transition'
% 11Jul2010: RoD editing of proofs comments for preprint; include bibliography in this file; replace png by pdf figures
%
%%%%%%%%%%%%%%%%%%%%%%%% Springer / AwR Book Team %%%%%%%%%%%%%%%%%%%%
\chapauthor{Roland Diehl\footnote{Max Planck Institut f\"ur extraterrestrische Physik, 85748 Garching, Germany}\footnote{DOI 10.1007/978-3-642-12698-7-1; Part of ``Astronomy with Radioactivities'', Eds. Roland Diehl, Dieter H. Hartmann, and Nikos Prantzos, Springer Lect. Notes in Physics, Vol. 218, 2010}
}
\chapter{Introduction to Astronomy with Radioactivity}
\label{intro}

\section{The Origin of Radioactivity}
\begin{figure}
  \includegraphics[width=\textwidth]{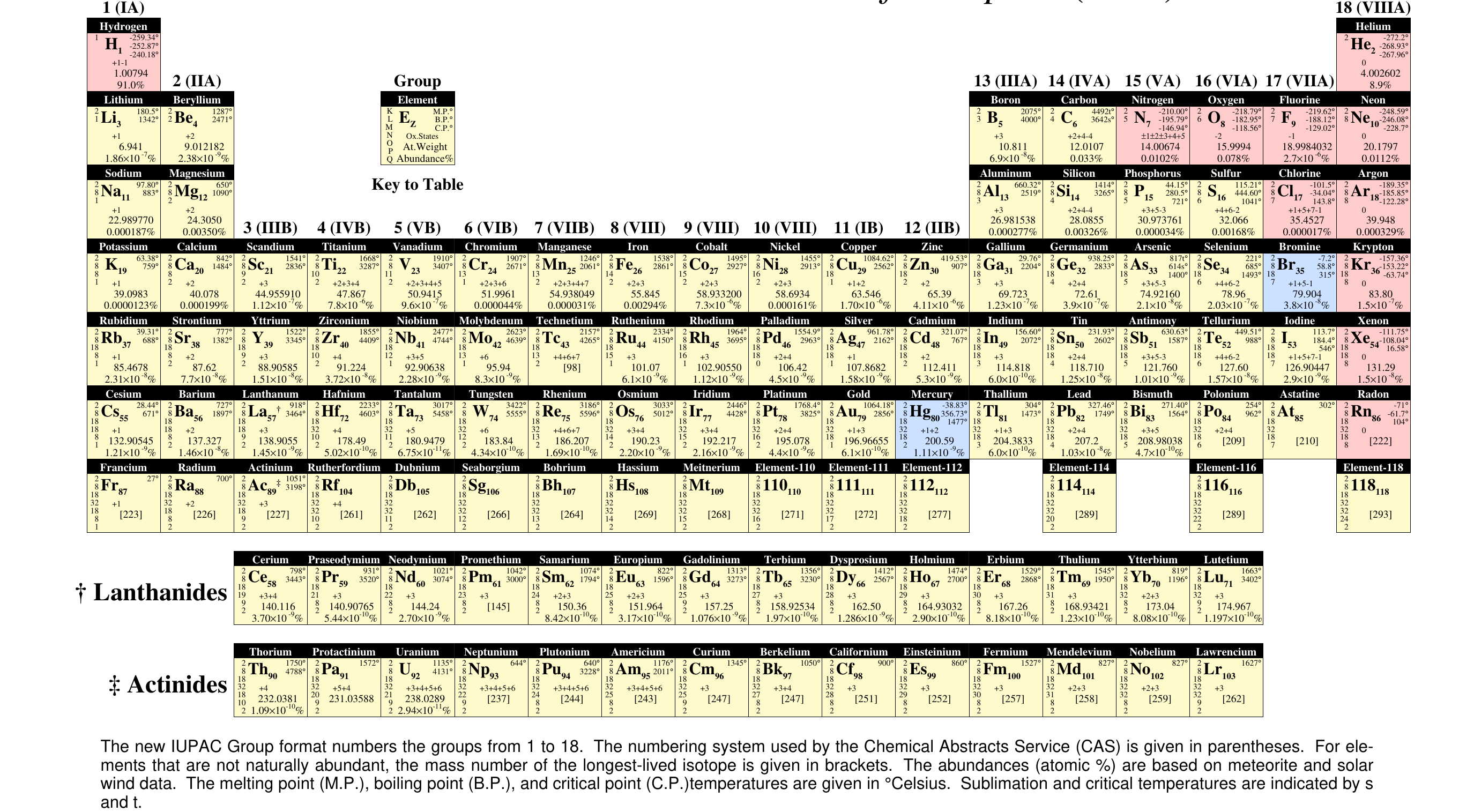}\\
  \caption{The periodic table of elements, grouping chemical elements according to their chemical-reaction properties and their atomic weight, after Mendeleyev (1869)}\label{fig_1_periodic_table}
\end{figure}

The nineteenth century spawned various efforts to bring order into the elements encountered in nature. Among the most important was an inventory of the {\it elements} \index{chemical elements} \index{elements!chemical} assembled by the Russian chemist Dimitri Mendeleyev \index{Mendeleyev, D.} in 1869, which grouped elements according to their chemical properties, their {\it valences}, as derived from the compounds they were able to form, at the same time sorting the elements by atomic weight. The genius of Mendeleyev lay in his confidence in these sorting principles, which enforce gaps in his table for expected but then unknown elements, and Mendeleyev was able to predict the physical and chemical properties of such elements-to-be-found. The tabular arrangement invented by Mendeleyev (Fig.~\ref{fig_1_periodic_table}) still is in use today, and is being populated at the high-mass end by the great experiments in heavy-ion collider laboratories to create the short-lived elements predicted to exist. The second half of the nineteenth century thus saw scientists being all-excited about chemistry and the fascinating discoveries one could make using Mendeleyev's sorting principles. Note that this was some 30~years before sub-atomic particles and the atom were discovered. Today the existence of 112 elements is firmly established, the (currently-heaviest) element no. 112 was officially named \emph{Copernicium (Cn)} in February 2010 by IUPAC, the international union of chemistry.

In the late nineteenth century, scientists also were excited about new types of penetrating radiation. Conrad R\"ontgen's \index{R\"ontgen, C.} discovery in 1895 of {\it X-rays} as a type of electromagnetic radiation is important for understanding the conditions under which Antoine Henri Becquerel \index{Becquerel, A.H.} discovered radioactivity in 1896. Becquerel also was engaged in chemical experiments, in his research on phosphorescence exploiting the chemistry of photographic-plate materials. At the time, Becquerel had prepared some plates treated with uranium-carrying minerals, but did not get around to make the planned experiment. When he found the plates in their dark storage some time later, he accidentally processed them, and was surprised to find an image of a coin which happened to have been stored with the plates. Excited about X-rays, he believed he had found yet another type of radiation. Within a few years, Becquerel with Marie and Pierre Curie \index{Curie, M.} and others recognized that the origin of the observed radiation were elemental transformations of the uranium minerals, and the physical process of {\it radioactivity} had been found. The revolutionary aspect of elements being able to spontaneously change their nature became masked at the beginning of the twentieth century, when sub-atomic particles and the atom were discovered. But it is worth emphasizing that well before atomic and quantum physics was known, the physics of weak interactions had been discovered in its form of {\it radioactivity}.

In an ensemble consisting of a large number of identical radioactive isotopes, it is well-known that the number remaining declines exponentially with time: 
Radioactive decay \index{decay!radioactive} is described by
\begin{equation}
 {{dN}\over{dt}} =  -\lambda \cdot N 
\end{equation}
Here $N$ is the number of isotopes, and the {\it radioactive-decay constant} $\lambda$ \index{decay!constant}is the inverse of the decay time \index{decay!time} $\tau$.
The decay time $\tau$ is the time after which the number of isotopes is reduced by decay to $1/e$ of the original number:

\begin{equation}
\label{eq_1} \index{decay!exponential}
 N = N_0 \cdot exp{-t\over\tau} 
\end{equation}

The radioactive half-life $T_{1/2}$, \index{decay!half life}correspondingly, is defined as the time after which the number of isotopes is reduced by decay to $1/2$ of the original amount, with
\begin{equation}
   T_{1/2} = {\tau \over ln(2)} 
\end{equation}

\begin{figure}
  \includegraphics[width=\textwidth]{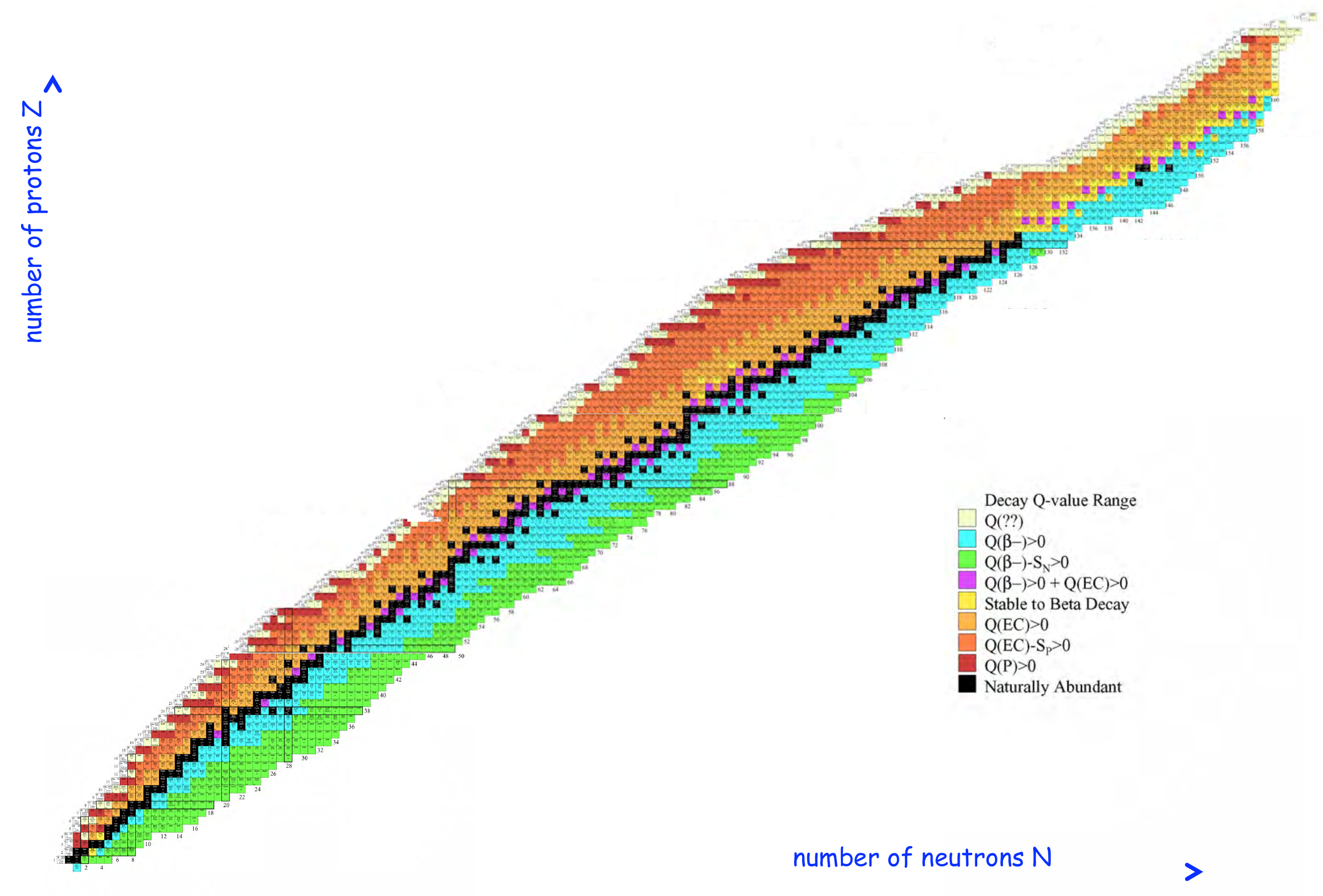}\\
  \caption{The table of isotopes, showing nuclei in a chart of neutron number (abscissa) versus proton number (ordinate). The stable elements are marked in black.
  All other isotopes are unstable, or radioactive, and will decay until a stable nucleus is obtained. }
  \label{fig_1_table-of-isotopes}
\end{figure}

\noindent Depending on the astrophysical objective, radioactive isotopes may be called \emph{short-lived} or \emph{long-lived} for identical lifetimes: The relation of the radioactive lifetime to astrophysical time scales of interest ist what matters. Examples are the utilization of \Al and \Fe ($\tau\sim$My) diagnostics of the early solar system (Chap.~6) or of nucleosynthesis source types (Chap.~3-5).

An {\it isotope} is defined by the number of its  two types of nucleons, {\it protons} (the number of protons defines the charge number Z) and {\it neutrons} (the sum of the numbers of protons and neutrons defines the mass number A), written as $^A$X for an element 'X'. Note that some isotopes may exist in different nuclear quantum states which have significant stability by themselves, so that transitions between these configurations may liberate the binding energy differences; such states of the same isotope are called {\it isomers}.

The above exponential decay law is a consequence of a surprisingly simple physical property: The probability per unit time for a single radioactive nucleus to decay is independent of the age of that nucleus. Unlike our commonsense experience with living things, decay does not become more likely as the nucleus ages. 
Radioactive decay is a nuclear transition from one set of nucleons constituting a nucleus to a different and energetically-favored set with the same number of nucleons. Different types of interactions can mediate such a transition (see below). In \emph{$\beta$-decays} it is mediated by the \emph{weak transition} \index{weak interaction} of a  neutron into a proton, or more generally, nucleons of one type into the other type\footnote{In a broader sense, nuclear physics may be considered to be similar to chemistry: elementary building blocks are rearranged to form different species, with macroscopically-emerging properties such as characteristic, well-defined energy releases in such transitions.}:
\begin{equation}\label{eq_n-decay}
n \longrightarrow p \mbox{ } + e^- \mbox{ } + \overline{\nu_e}
\end{equation}
If such a process occurs inside an atomic nucleus, the quantum state of the nucleus is altered. Depending on the variety of configurations in which this new state may be realized (i.e. the \emph{phase space} available to the decaying nucleus), this change may be more or less likely, in nature's attempt to minimize the total energy of a composite system of nucleons. 
The decay probability $\lambda$ per unit time for a single radioactive nucleus is therefore a property which is specific to each particular type of isotope. It can be estimated by Fermi's \emph{Golden Rule} formula \index{Golden Rule} though time-dependent perturbation theory \citep[e.g.][]{1962qume.book.....M}. When schematically simplified to convey the main ingredients, the decay probability is:
\begin{equation}\label{eq1}
\lambda = \frac{4\pi^2}{h} \mbox{ } V_{fi}^2 \mbox{ } \rho(W) = \frac{1}{\tau}
\end{equation}
\noindent where $\rho(W)$ is the number of final states having suitable energy $W$. The full theory involves an integral over the final kinematic states, suppressed here for simplicity. The matrix element $V_{fi}$ is the result of the transition-causing potential between initial and final states.  The parameter $\tau$, defined as 1/$\lambda$, can easily be shown to be the mean lifetime of the nucleus against radioactive decay.

In general, transitions occur from the ground state of the parent nucleus to an excited state of the daughter nucleus. But quantum mechanical transition rules may allow and even prefer other initial and final states. Excess binding energy will be transferred to the end products, which are the daughter nucleus, but also emitted (or absorbed, in the case of electron capture transitions) through leptons (electrons, positrons, neutrinos) and $\gamma$-ray photons. 

The occupancy of states is mediated by the \emph{thermal} excitation spectrum of the \emph{Boltzmann distribution} \index{Boltzmann distribution} of particles, populating states at different energies according to:
\begin{equation}
{dN \over dE} = G_j \cdot e^{-{{E}\over{k_BT}}} 
\end{equation}
\noindent Here $k_B$ is Boltzmann's constant, $T$ the temperature of the particle population, $E$ the energy, and $G_j$ the statistical weight factor of all different possible states $j$ which correspond to a specific energy $E$\footnote{States may differ in their quantum numbers, such as spin, or orbital-momenta projections; if they obtain the same energy $E$, they are called \emph{degenerate}.}.
In natural environments, particles will populate different states as temperature dictates. Transition rates among states thus will depend on temperature. Inside stars, and more so in explosive environments, temperatures can reach ranges which are typical for nuclear energy-level differences. Therefore, in cosmic sites, radioactive decay constants may be significantly different from what we measure in terrestrial laboratories on \emph{cold} samples (see Section~\ref{sec:1_processes} for more detail).

The atomic-shell environment of a nucleus may modify radioactive decay, in particular if a decay involves \emph{capture of an electron} \index{decay!electron capture} to transform a proton into a neutron. Such decays are inhibited in fully-ionized plasma conditions, due to the non-availability of electrons.
Which radioactive decays are to be expected, in a mix of nuclei such as could be imagined to exist in a thermal bath of nuclei? What are stable configurations of nucleons inside a nucleus? This involves an understanding of the nuclear forces, an area of current research and far from being fully understood. Nevertheless, a few general ideas appear well established. These can be summarized in the expression for nuclear masses \index{mass!nuclear} \citep{1935ZPhy...96..431W}:

\begin{equation}
 m(Z,A) = Z M_p + (A-Z) m_n - BE 
\end{equation}
with
\begin{equation}
 BE = a_{volume} A - a_{surface} A^{2/3} - a_{coulomb} {Z^2 \over {A^{1/3}}} - a_{asymmetry} {{{(a-2Z)}^2} \over {4A}} - {\delta \over A^{1/2}} 
\end{equation}

This description emphasizes that the total \emph{binding energy} (BE) \index{binding energy} is another free parameter for a system of nucleons, allowing them to adopt bound states of lower energy than the sum of the free nucleons. Thus, in a thermal mixture of nucleons, bound nuclei will be formed, and their abundance depends on their composition and shape, and on the overall system temperature, defining how the totally-available phase space of internal and kinetic energy states is populated.

The key gross features characterizing the bound systems with local maxima of binding energy are (1) the \emph{odd-even} \index{odd-even effect} effect described by the last term, which results in odd-nucleon nuclei being less favored that even-nucleon nuclei, and (2) the neutron excess favored by the asymmetry term, which results in heavier nuclei being relatively more neutron rich.

As an illustration of the available nuclear configurations, \index{isotopes!table of} Fig.~\ref{fig_1_table-of-isotopes} shows the stable (black) and unstable isotopes, the latter decaying by $\beta^-$-decay (blue) and $\beta^+$-decay (orange).
The \emph{ragged} structure signifies that there are systematic variations of nuclear stability with nucleon number, some nucleonic numbers allowing for a greater variety of stable configurations of higher binding energy. These are, in particular, {\it magic numbers} of protons and neutrons of 2, 8, 20, 28, 50, and 82.
We now know approximately 3100 such \emph{isotopes} making up the 114 now-known \index{elements!known} chemical elements, but only 286 of these isotopes are considered stable. (The latest (7$^{th}$) edition of the Karlsruher Nuklidkarte \citep{2007KNucChart..7} lists 2962 experimentally-observed isotopes and 652 isomers, its first edition (1958) included 1297 known isotopes of 102 then-known \index{elements!superheavy} elements. At the time of writing, elements 115 and 113 are the most massive superheavy elements which have been synthesized and found to exist at least for short time intervals, although more massive elements may exist in an island of stability beyond).  

Unstable isotopes, once produced, will be \emph{radioactive}, i.e. they will transmute to other isotopes through nuclear interactions, until at the end of such a decay chain a stable isotope is produced. Weak interactions will mediate transitions between protons and neutrons and lead to neutrino emission, involvements of atomic-shell electrons will result in X-rays from atomic-shell transitions after electron capture and internal-conversion transitions, and $\gamma$-rays will be emitted in electromagnetic transitions between excitation levels of a nucleus. 

It is the subject of this book to explain in detail the astrophysical implications of this characteristic process of nuclear rearrangements, and what can be learned from measurements of the messengers of radioactive decays.
But first we describe the phenomenon of radioactivity in more detail.

%%%% Insert from DDC %%%%%%%
\section{The Processes of Radioactivity}
\label{sec:1_processes}
  After Becquerel's discovery of radioactivity \index{radioactivity}  in 1896, Rutherford \index{Rutherford, E.} and others found out in the early 20$^{\rm{th}}$ century that there were different types of radioactive decay \citep{1903PPSL...18..595R}. They called them \emph{$\alpha$ decay, $\beta$ decay} and \emph{$\gamma$ decay}, terms which are still used today. It was soon understood that they are different types of interactions, all causing the same, spontaneous, and time-independent decay of an unstable nucleus into another and more stable nucleus. 

 \noindent {\it Alpha decay}\index{alpha decay}\index{decay!alpha decay}  : This describes the ejection of a $^4$He nucleus from the parent radioactive nucleus upon decay. $^4$He nuclei have since been known also as \emph{alpha particles} for that reason. This decay is intrinsically fast, as it is caused by the \emph{strong} nuclear interaction quickly clustering the nucleus into an alpha particle and the daughter nucleus. Since $\alpha$-nuclei are tighly-bound, they have been found as sub-structures even within nuclei. In the cases of nuclei much heavier than Fe, a nucleus thus consisting of many nucleons and embedded $\alpha$ clusters can find a preferred state for its number of nucleons by separation of such an $\alpha$ cluster, liberating the binding-energy difference\footnote{The binding energy \emph{per nucleon} is maximized for nucleons bound as a Fe nucleus.}. In such heavy nuclei, Coulomb repulsion helps to overcome the potential barrier which is set up by the strong nuclear force, and decay can occur through emission of an $\alpha$ particle. The $\alpha$ particle \emph{tunnels}, with some calculable probability, through the potential barrier, towards an overall more stable and less-energetic assembly of the nucleons.

An example of $\alpha$ decay is $_{88}$Ra$^{226}$ $\Rightarrow$ $_{86}$Rn$^{222}$ + $_2$He$^4$, which is one step in the decay series starting from $^{238}$U. The daughter nucleus , $_{86}$Rn$^{222}$, has charge $Z-2$, where $Z$ is the original charge of the radioactive nucleus ($Z$=88 in this example), because the $\alpha$ particle carried away two charge units from the original radioactive nucleus. Such decay  frequently leads to an excited state of the daughter nucleus. Kinetic energy $E_{\alpha}$ for the $\alpha$ particle is made available from the nuclear binding energy liberation expressed by the \emph{Q-value} of the reaction \index{Q value} if the mass of the radioactive nucleus exceeds the sum of the masses of the daughter nucleus and of the helium nucleus\footnote{These masses may be either nuclear masses or atomic masses, the electron number is conserved, and their binding energies are negligible, in comparison.}:
\begin{equation} \index{isotopes!226Ra}
Q_{\alpha} = [M(_{88}\rm{Ra}^{226}) - M(_{86}\rm{Rn}^{222}) - M(_2\rm{He}^4)]\rm{c}^2
\end{equation}
The range of the $\alpha$ particle (its stopping length) is about 2.7 cm in standard air (for an $\alpha$ particle with E$_{\alpha}$ of 4 MeV), and it will produce about 2$\times$10$^5$ ionizations before being stopped. Even in a molecular cloud, though  its range would be perhaps 10$^{14}$ times larger, the $\alpha$ particle would not escape from the cloud. Within small solids (dust grains), the trapping of radioactive energy from $\alpha$ decay provides a source of heat which may result in characteristic melting signatures\footnote{Within an FeNi meteorite, e.g., an $\alpha$ particle from radioactivity has a range of only $\sim$10~$\mu$m.}.
% Alpha particles are relatively easy to stop. Such details were engagingly described by Rutherford, Chadwick \& Ellis in a 1930 classic book \citep{Ruth30}. \\

\noindent {\it Beta decay:}\index{beta decay} \index{decay!beta decay}  This is the most-peculiar radioactive decay type, as it is caused by the nuclear \emph{weak interaction} which converts neutrons into protons and vice versa. The neutrino $\nu$ \index{neutrino} carries energy and momentum to balance the dynamic quantities, as Pauli famously proposed in 1930 (Pauli did not publish this conjecture until 1961 in a letter he wrote to colleagues). The $\nu$ was given its name by \index{Fermi, E.} Fermi, and was discovered experimentally in 1932 by \index{Chadwick, J.} James Chadwick, i.e. \emph{after} Wolfgang Pauli \index{Pauli, W.} had predicted its existence. Neutrinos from the Sun have been discovered to \emph{oscillate} between flavors. $\beta$ decays are being studied in great detail by modern physics experiments, to understand the nature and mass of the $\nu$. Understanding $\beta$ decay challenges our mind, as it involves several such unfamiliar concepts and particles.

There are three types\footnote{We ignore here two additional $\beta$ decays which are possible from $\nu$ and $\overline{\nu}$ captures, due to their small probabilities.} of $\beta$-decay: 
\begin{equation}\label{eq_beta+-decay}
^A_ZX_N\mbox{ }  \longrightarrow \mbox{ } ^A_{Z-1}X_{N+1} \mbox{ } + e^+ \mbox{ } + \nu_e
\end{equation}
\begin{equation}\label{eq_beta--decay}
^A_ZX_N \mbox{ } \longrightarrow \mbox{ } ^A_{Z+1}X_{N-1} \mbox{ } + e^- \mbox{ } + \overline{\nu_e}
\end{equation}
\begin{equation}\label{eq_beta--decay}
^A_ZX_N \mbox{ } + e^- \longrightarrow \mbox{ } ^A_{Z-1}X_{N+1}  \mbox{ } + {\nu_e}
\end{equation}
In addition to eq.~\ref{eq_n-decay} (\emph{$\beta^-$~decay}), these are the conversion of a proton into a neutron (\emph{$\beta^+$~decay}), and \emph{electron capture}.
The weak interaction itself involves two different aspects with intrinsic and different strength, the  vector and axial-vector couplings. The $V_{fi}^2$ term in eq.~\ref{eq1} thus is composed of two terms. These result in  \index{Fermi!transition}\index{Gamow Teller transition} \emph{Fermi} and \emph{Gamow-Teller transitions}, respectively \citep[see][for a review of weak-interaction physics in nuclear astrophysics]{2003RvMP...75..819L}.

An example of $\beta$ decay is \index{isotopes!13N} $_7^{13}$N $\longrightarrow \mbox{ }_6^{13}$C~+~e$^{+}$~$+$~$\nu$, having mean lifetime $\tau$ near 10 minutes. The kinetic energy $Q$ of the two leptons, as well as the created electron's mass, must be provided by the radioactive nucleus having greater mass than the sum of the masses of the daughter nucleus and of an electron (neglecting the comparatively-small neutrino mass).
\begin{equation}
Q_{\beta} =[M(_7^{13}\rm{N}) - M(_6^{13}\rm{C})- m_{e}]c^2
\end{equation}
\noindent where these masses are nuclear masses, not atomic masses. A small fraction of the energy release $Q_{\beta}$ appears as the recoil kinetic energy of the daughter nucleus, but the remainder appears as the kinetic energy of electron and of neutrino.

Capture of an electron is a \emph{two-particle} reaction, the bound atomic electron $e^{-}$ or a free electron in hot plasma being required for this type of $\beta$ decay. Therefore, depending on availability of the electron, electron-capture $\beta$ decay lifetimes can be very different for different environments. 
	In the laboratory case, electron capture usually involves the 1s electrons of the atomic structure surrounding the radioactive nucleus, because those present their largest density at the nucleus. 
	
	In many cases the electron capture competes with $e^{+}$ $+$ $\nu$ emission. In above example, $^{13}$N can decay not only by emitting $e^{+}$ $+$ $\nu$, but also by capturing an electron: $_7^{13}$N~+~e$^{-}\longrightarrow_6^{13}$C~+~$\nu$. In this case the capture of a 1s electron happens to be much slower than the rate of e$^+$ emission. But cases exist for which the mass excess is not large enough to provide for the creation of the $e^{+}$ mass for emission, so that only electron capture remains to the unstable nucleus to decay. Another relevant example is the decay of $^7$Be. Its mass excess over the daugther nucleus $^7$Li is only 0.351 MeV. This excess is insufficient to provide for creation of the rest mass of an emitted $e^{+}$, which is 0.511 MeV. Therefore, the $^7$Be nucleus \index{isotopes!7Be} is stable against $e^{+}$ $+$ $\nu$ emission. However, electron capture adds 0.511 MeV of rest-mass energy to the mass of the $^7$Be nucleus, giving a total 0.862 MeV of energy above the mass of the $^7$Li nucleus. Therefore, the $e^{-}$ capture process (above) emits a monoenergetic neutrino having $E_{\nu}$= 0.862 MeV\footnote{This neutrino line has just recently been detected by the Borexino collaboration arriving from the center of the Sun \citep{2008PhRvL.101i1302A}.}.

	The situation for electron capture processes differs significantly in the interiors of stars and supernovae: Nuclei are ionized in plasma at such high temperature. The capture lifetime of $^7$Be, for example, which is 53 days against 1s electron capture in the laboratory, is lengthened to about 4 months at the solar center \citep[see theory by][]{1964ApJ...139..318B,1983NuPhA.404..578T}, where the free electron density is less at the nucleus.

	The range of the $\beta$ particle (its stopping length) is small, being a charged particle which undergoes Coulomb scattering. An MeV electron has a range of several meters in standard air, during which it loses energy by ionizations and inelastic scattering. 
	%Trapping in solids is not an issue with electrons, but whether $e^{+}$ particles 
Energy deposit or escape is a major issue in expanding envelopes of stellar explosions, supernovae (positrons \index{positron} from $^{56}$Co \index{isotopes!56Co}\index{isotopes!44Ti} and $^{44}$Ti) and novae (many $\beta^+$ decays such as $^{13}$N) (see Chapters~4, 5, and~7 for a discussion of the various astrophysical implications). 
Even in small solids and dust grains, energy deposition from \Al  $\beta$-decay, for example, injects 0.355~W~kg$^{-1}$ of heat. This is sufficient to result in melting signatures, which have been used to study condensation sequences of solids in the early solar system (see Chapter~6).
	%The nuclear recoil during beta decay is so low that the daughter nucleus almost always remains within the solid in which the decay occurs, even if it is only microns in size. So a daughter $^{129}$Xe atom created by radioactive $^{129}$I beta decay is hardly dislodged from its chemical site. The range of the emitted neutrino is of course very large, even escaping from the center of the sun.

\begin{figure}
  \includegraphics[width=\textwidth]{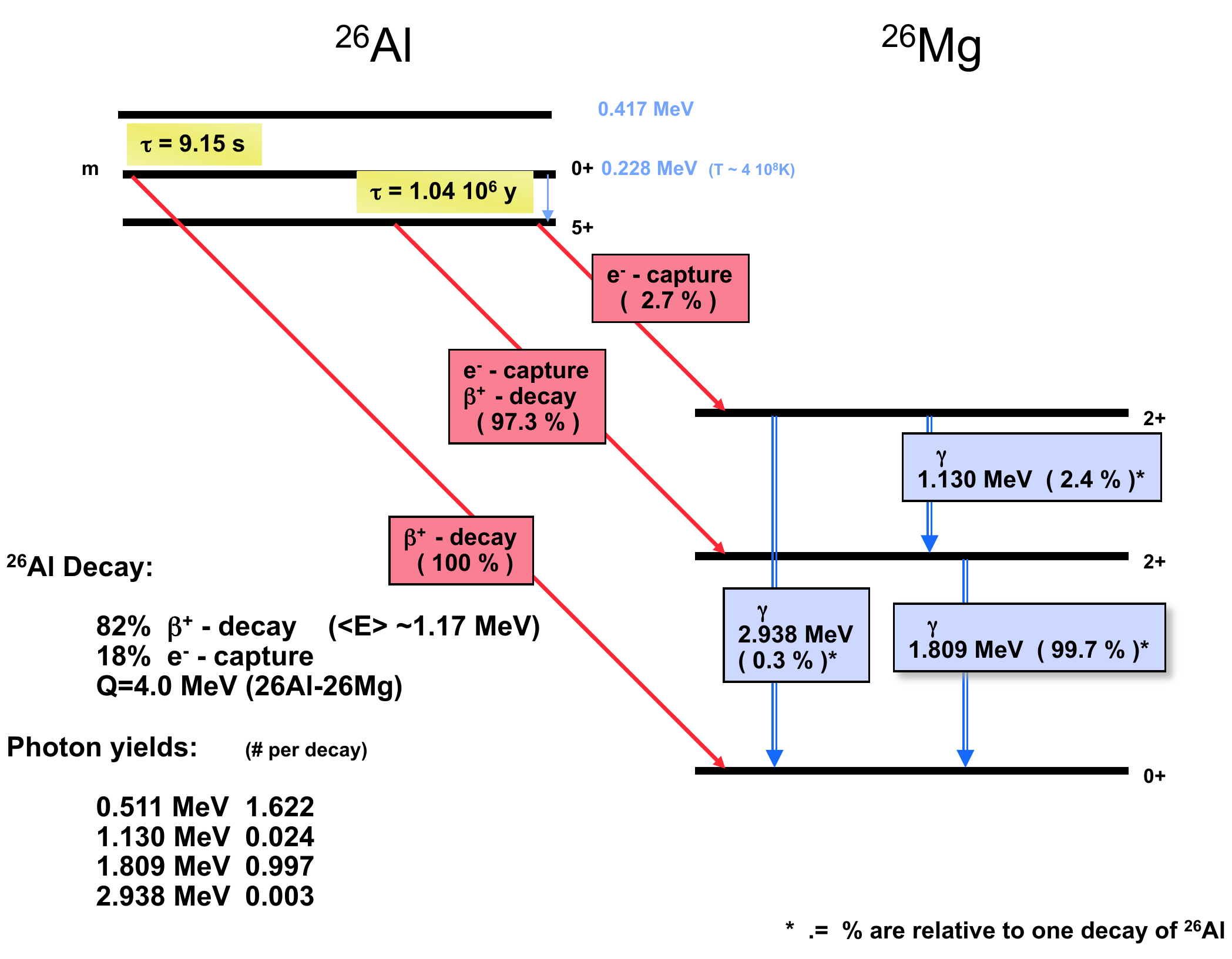}
  \caption{\Al decay. The \Al nucleus ground state has a long radioactive lifetime, due to the large spin difference of its state to lower-lying states of the daughter nucleus $^{26}$Mg. \index{isotopes!26Al} An isomeric excited state of \Al exists at 228 keV excitation energy. If thermally excited, \Al may decay through this state. Secondary products, lifetime, and radioactive energy available for deposits and observation depend on the environment. }
  \label{fig_1_26Al-decay}
\end{figure}

\noindent {\it Gamma decay:}\index{gamma decay} \index{decay!gamma decay}   In $\gamma$ decay the radioactive transition to a different and more stable nucleus is mediated by the \emph{electromagnetic interaction}. A nucleus relaxes from its excited configuration of the nucleons to a lower-lying state of the same nucleons. This is intrinsically a fast process; typical lifetimes for excited states of an atomic nucleus are 10$^{-9}$seconds. We denote such electromagnetic transitions of an excited nucleus \emph{radioactive $\gamma$-decay} when the decay time of the excited nucleus is considerably longer and that nucleus thus may be considered a temporarily-stable configuration of its own, a \emph{metastable} nucleus. 

How is stability, or instability, of a nuclear-excited state effected?
Electromagnetic transitions 
\begin{equation}\label{eq_photon-decay}
A^{\star} \longrightarrow A^{g.s.} + \gamma
\end{equation}
must satisfy spin (angular momentum) conservation in the transition. The spin state of a nuclear state is a property of the nucleus, and reflects how protons and neutrons are spread over the quantum-mechanically allowed \emph{shells} or nucleon wave functions (as expressed in the \emph{shell model} view of an atomic nucleus).
The photon (\emph{$\gamma$ quantum}) emitted (eq.\ref{eq_photon-decay}) will thus have a \emph{multipolarity} \index{multipolarity} dictated by the initial and final states of the nucleus. Dipole radiation is most common and has multipolarity 1, emitted when initial and final state have angular momentum difference $\Delta l=1$. Quadrupole radiation (multipolarity 2, from $\Delta l=2$) is $\sim$6 orders of magnitude more difficult to obtain, and likewise, higher multipolarity transitions are becoming less likely by the similar probability decreases 
(the \emph{Weisskopf estimates} \index{Weisskopf estimates} \citep[see][]{1951PhRv...83.1073W}). 
This explains why some excited states in atomic nuclei are much more long-lived (\emph{meta-stable}) \index{metastable} than others; their transitions to the ground state are considered as \emph{radioactivity} and called \emph{$\gamma$ decay}.

 The range of a $\gamma$-ray (its stopping length) is typically about 5-10~g~cm$^{-2}$ in passing through matter of all types. Hence, except for dense stars and their explosions, radioactive energy from $\gamma$ decay is of astronomical implication only\footnote{Gamma-rays from nuclear transitions following $^{56}$Ni decay \index{isotopes!56Ni} (though this is a $\beta$ decay by itself) inject radioactive energy through $\gamma$-rays from such nuclear transitions into the supernova \index{supernova!light curve} envelope, where it is absorbed in scattering collisions and thermalized. This heats the envelope such that thermal and optically bright supernova light is created. Deposition of $\gamma$-rays from nuclear transitions are the engines which make supernovae to be bright light sources out to the distant universe, used in cosmological studies \citep{2000A&ARv..10..179L} to, e.g., support evidence for \emph{dark energy}.}. 
 %If located within small solids, the radioactive energy released in $\gamma$ decay can be sufficient to heat and melt material, leaving behind characteristic morphology of the solid. %\Al included in small solids of the early solar system (this is beta decay!)

An illustrative example of radioactive decay is the \Al nucleus. \index{decay!26Al} \index{isotopes!26Al}\index{isotopes!26Mg} Its decay scheme is illustrated in Fig.~\ref{fig_1_26Al-decay}. The ground state of \Al is a $5+$ state. Lower-lying states of the neighboring isotope $^{26}$Mg have states $2+$ and $0+$, so that a rather large change of angular momentum $\Delta l$ must be carried by radioactive-decay secondaries. This explains the large $\beta$-decay lifetime of \Al of $\tau\sim$1.04~10$^6$~y.  
In the level scheme of \Al, excited states exist at energies 228, 417, and 1058~keV. The $0+$ and $3+$ states of these next excited states are more favorable for decay due to their smaller angular momentum differences to the $^{26}$Mg states, although $\Delta l=0$ would not be \emph{allowed} for the 228~keV state to decay to $^{26}$Mg's ground state. This explains its relatively long lifetime of 9.15~s, and it is a \emph{metastable} \index{metastable} state of \Al.  If thermally excited, which would occur in nucleosynthesis sites exceeding a few 10$^8$K, \Al may decay through this state without $\gamma$-ray emission as $^{26}\rm{Al}^{g.s.} + \gamma \longrightarrow ^{26}\rm{Al}^{m} \longrightarrow ^{26}\rm{Mg} + e^+$, while the ground state decay is predominantly a \emph{$\beta^+$ decay} through excited $^{26}$Mg states and thus including $\gamma$-ray emission. Secondary products, lifetime, and radioactive energy available for deposits and observation depend on the environment.

%%%%%%%%%%%%%%%%%%%%%%%%%%%%%%%%%%%%%%%%%%%%%%%%%%%%%%%%%%%%%%%%%
\section{Radioactivity and Cosmic Nucleosynthesis}

\begin{figure}
  \includegraphics[width=\textwidth]{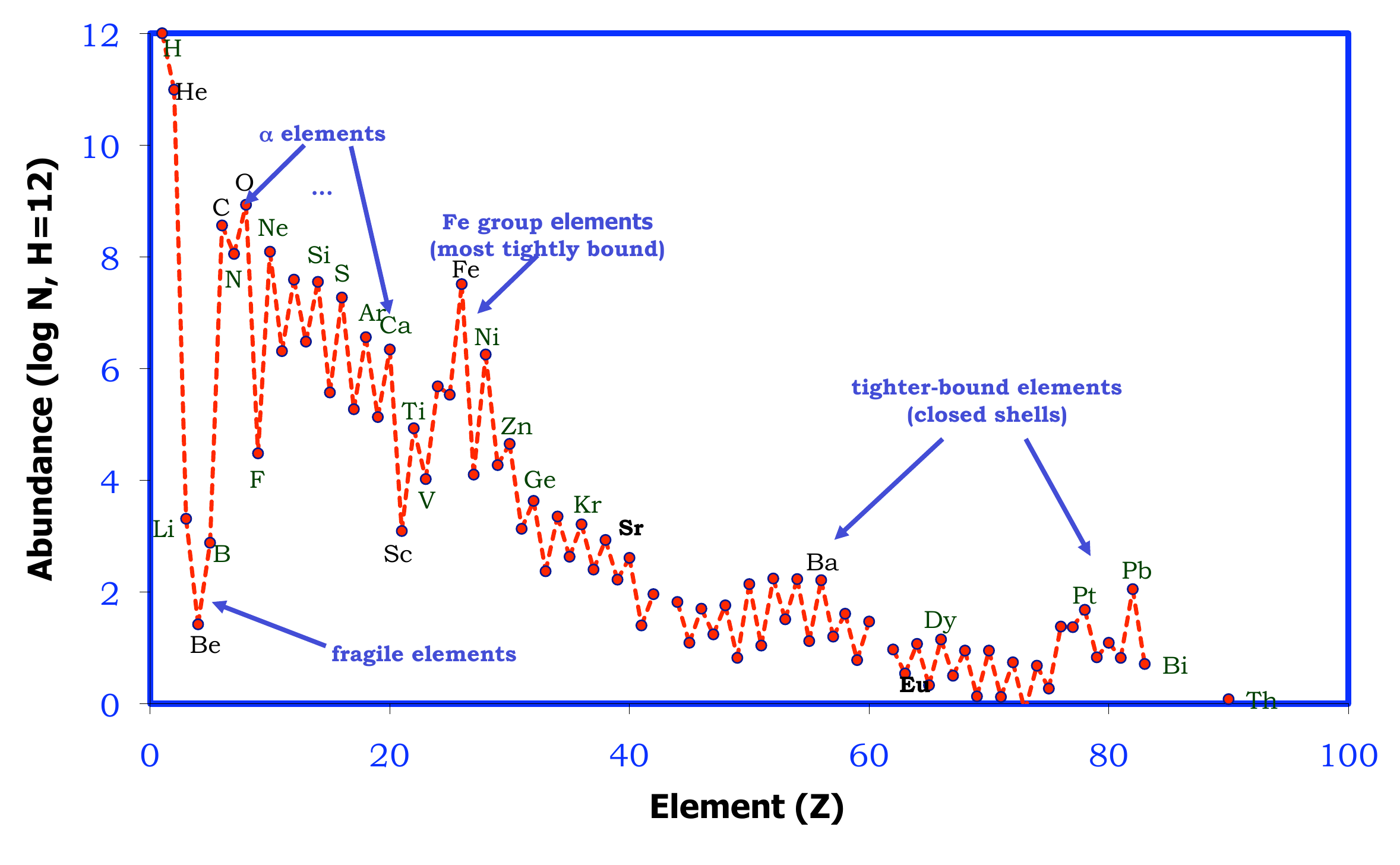}
  \caption{The abundance of elements \index{abundances!cosmic} \index{abundances!elemental}     in the present-day nearby universe. Abundances (by number) are shown in a logarithmic scale, and span 12 orders of magnitude. The interplay of nuclear properties (several are indicated in the graph) with environmental conditions in cosmic nucleosynthesis sites has created this complex abundance pattern during the course of cosmic history.}
  \label{fig_1_abundances}
\end{figure}

Nuclear reactions in cosmic sites re-arrange the basic constituents of atomic nuclei (neutrons and protons) among the different configurations which are allowed by Nature.
Throughout cosmic history, such reactions occur in different sites, and lead to rearrangements of the relative abundances of cosmic nuclei, a process called \emph{cosmic chemical evolution}\index{chemical evolution!cosmic}. \footnote{We point out that there is no chemistry involved; the term refers to changes in abundances of chemical elements, which are a result of the changes in abundances of isotopes.}

The \emph{cosmic abundance} of a specific isotope \index{abundances!normalizations} is expressed in different ways, depending on the purpose. Counting the atoms of isotope $i$ per unit volume, one obtains $n_i$, the number density of atoms of species $i$ (atoms~cm$^{-3}$).  The interest of cosmic evolution and nucleosynthesis lies in the fractional abundances of species $i$ related to the total, and how it is altered by cosmic nuclear reactions. Observers count a species $i$ and relate it to the abundance of a reference species. For \index{abundances!astronomical} astronomers this is hydrogen. Hydrogen is the most abundant element throughout the universe, and easily observed through its characteristic atomic transitions in spectroscopic astronomical measurements. 
Using the definition of  Avogadro's constant $A_{Av}$  as the number of atoms which make up $A$ grams of species $i$ (i.e., one mole), we can obtain abundances \emph{by mass}; $A_{Av}=6.02214~10^{23}$ atoms~mole$^{-1}$. 
The mass contained in a particular species $S$ results from scaling its abundance $N_S$ by its atomic weight $A$. 
%The abundance scale is set to $log(N_H)=X_H=12$ (see Fig.~\ref{fig_1_abundances}). Mineralogists and \index{abundances!meteoritic} meteoritic scientists use Silicon as their reference species and set $log(N_{Si})=X_{Si}=6$. 
We can get a global measure for cosmic evolution of the composition of matter by tracing how much of the total mass is contained in hydrogen, helium, and the remainder of elements called \emph{metals}\footnote{This nomenclature may be misleading, it is used by  convenience among astrophysicists. Only a part of these elements are actually metals.}, calling these quantities $X$ for hydrogen abundance, $Y$ for helium abundance, and $Z$ for the cumulative abundance of all nuclei heavier than helium. We call these \emph{mass fractions} of hydrogen $X$, helium $Y$, and metals $Z$, with $X+Y+Z=1$.  
The metalicity $Z$ is a key parameter used to characterize the evolution of elemental and isotopic composition of cosmic matter. The astronomical abundance scale is set from most-abundant cosmic element Hydrogen to $log(X_H)=12$ (Fig.~\ref{fig_1_abundances}), but mineralogists and meteoriticians use $Si$ as their reference element and set $log(X_{Si})=6$. 

We often relate abundances also to our best-known reference, the solar system, denoting \emph{solar-system} values by the $\odot$ symbol. Abundances of a species $S$ are then expressed in \index{abundances!bracket notation} \emph{bracket notation}\footnote{Deviations from the standard may be small, so that $\left[\frac{[S_1}{S_2}\right]$ may be expressed in $\delta$ units (parts per mil), or $\epsilon$ units (parts in 10$^4$), or ppm and ppb; $\delta(^{29}Si/^{28}Si)$ thus denotes excess of the $^{29}$Si/$^{28}$Si isotopic ratio above solar values in units of 0.1\%.}
 as \index{abundances!cosmic} \index{abundances!solar} \index{abundances!standard} 
\begin{equation}
\left[ \frac{ S }{ H } \right] \equiv log (\frac{X_S}{X_H})_{\star}  -  log (\frac{X_S}{X_H})_{\odot}  
%\frac{\lbrack S \rbrack}{\lbrack H \rbrack}  \equiv log (\frac{N_S}{N_H})_{\star}  -  log (\frac{N_S}{N_H})_{\odot}  
%\equiv log \epsilon_{S,\star} - log \epsilon_{S,\odot}  \equiv \delta_S
\end{equation}
\noindent
Depending on observational method and precision, our astronomical data are  \emph{metalicity}, elemental \emph{enrichments} with respect to solar abundances, or isotopic abundances. Relations to nuclear reactions are therefore often indirect. Understanding the nuclear processing of matter in the universe is a formidable challenge, often listed as one of the \emph{big questions} of science.

After nucleosynthesis during the Big Bang, \index{abundances!big bang} most-abundant were hydrogen (protons) and helium; the total amount of nuclei heavier than He (the \emph{metals}) was less than 10$^{-9}$ (by number, relative to hydrogen \citep{2007ARNPS..57..463S}). Today, the total mass fraction of metals in {\it solar abundances}, our local reference which seems remarkably universal, is $Z=0.0134$  \citep{2009ARA&A..47..481A}, compared to a hydrogen mass fraction of\footnote{This implies a \emph{metalicity} of solar matter of 1.4\%. Earlier than $\sim$2005, the commonly-used value for solar metalicity had been 2\%.} $X=0.7381$. This growth by about seven orders of magnitude is the effect of cosmic nucleosynthesis. Nuclear reactions in stars, supernovae, novae, and other places where nuclear reactions may occur, all contribute. But it also is essential that the nuclear-reaction products will be made available to observable cosmic gas and solids eventually. This book will also discuss our observational potential for cosmic isotopes, and we address the constraints and biases which limit our ability to draw far reaching conclusions.

The growth of isotopic and elemental abundances from cosmic nucleosynthesis does not occur homogeneously. Rather, the cosmic abundances observed today span a dynamic range of twelve orders of magnitude between abundant hydrogen and rare heavy elements (Fig.~\ref{fig_1_abundances}). Moreover, the elemental abundance pattern already illustrates clearly the prominent effects of nuclear structure (see Fig.~\ref{fig_1_abundances}): Iron elements are among the most-tightly bound nuclei, and locally elements with even numbers of nucleons are more tightly bound than elements with odd numbers of nuclei. The Helium nucleus (\emph{$\alpha$-particle}) also is more tightly bound than its neighbors in the chart of nuclei, hence all elements which are multiples of ($\alpha$'s are more abundant than their neighbors. 

Towards the heavier elements beyond the Fe group, abundances drop by about five orders of magnitude again, signifying a substantially-different production process than the mix of charged-particle nuclear reactions that produced the lighter elements: \emph{neutron capture} on Fe \emph{seed nuclei}. The two abundance peaks seen for heavier elements are the results of different environments for cosmic neutron capture reactions (the \emph{r-process} and \emph{s-process}), both determined by neutron capture probabilities having local extrema near \emph{magic numbers}. The different peaks arise from the particular locations at which the processes' reaction path encounters these \emph{magic nuclei}, as neutron captures proceed much faster (slower) than beta decays in the $r$~process ($s$~process)..

The issues in cosmic nucleosynthesis research are complex, and cover the astrophysics of stars, stellar explosions, nuclear reactions on surfaces of compact stars and in interstellar space. For each of the potential nuclear-reaction sites, we need to understand first how nuclear reactions proceed under the local conditions, and then how material may be ejected into interstellar space from such a source. None of the nucleosynthesis sites is currently understood to a sufficient level of detail which would allow us to sit back and consider cosmic nucleosynthesis \emph{understood}. For example, for the Sun, where one would assume we know most detail, solar neutrino \index{neutrino} measurements have been a puzzle only alleviated in recent years with the revolutionary adoption of non-zero masses for neutrinos. 
Even if we consider this a sufficient explanation, solar elemental abundances have recently been revised by almost a factor two based on consistency arguments between three-dimensional models of the solar photosphere and measured elemental line profiles from this photosphere; but helio-seismological measurements for the Sun are in significant conflict with the underlying model which describes the interior structure of the Sun in these three-dimensional models \citep{2009ARA&A..47..481A}. As another example, there are two types of supernova explosions, core-collapse supernovae marking the final gravitational collapse of a massive star once its nuclear fuel is exhausted, and thermonuclear supernovae, thought to originate from detonation of degenerate stars once they exceed a critical threshold for nuclear burning of Carbon. For neither of these supernovae, a \emph{physical} model is available, which would allow us to calculate and predict the outcome (energy and nuclear ashes) from such an explosion under given, realistic, initial conditions (see Ch. 4 and 5). Much research remains to be done in cosmic nucleosynthesis.

One may consider measurements of cosmic material in all forms to provide a wealth of data, which now has been exploited to understand cosmic nucleosynthesis. Note, however, that cosmic material as observed has gone through a long and ill-determined journey. We need to understand the trajectory in time and space of the progenitors of our observed cosmic-material sample if we want to interpret it in terms of cosmic nucleosynthesis. 
This is a formidable task, necessary for distant cosmic objects, but here averaging assumptions help to simplify studies. For more nearby cosmic objects where detailed data are obtained, astrophysical models quickly become very complex, and also need simplifying assumptions to operate for what they are needed. It is one of the objectives of cosmic nucleosynthesis studies to contribute to proper models for processes in such evolution, which are sufficiently isolated to allow their separate treatment. Nevertheless, carrying out \emph{well-defined experiments} for a source of cosmic nucleosynthesis remains a challenge, due to this often ill-constrained history (see Ch. 6 and 7).

The special role of radioactivity in such studies is contributed by the intrinsic decay of such material after it has been produced in cosmic sites. This brings in a new aspect, the clock of the radioactive decay. With such additional information, changes in isotopic abundances with time will occur naturally, and will leave their imprints in observable isotopic abundance records. For example, the observation of unstable technetium in stellar atmospheres of AGB stars was undisputable proof of synthesis of this element inside the same star, because the evolutionary time of the star exceeds the radioactive lifetime of technetium. Another example, observing radioactive decay $\gamma$-ray lines from short-lived Ni isotopes from a supernova is clear proof of its synthesis in such explosions; measuring its abundance through $\gamma$-ray brightness is a prominent goal for direct \emph{calibration} of processes in the supernova interior. A last example, solar-system meteorites \index{abundances!solar} show enrichments in daughter products of characteristic radioactive decays, such as $^{26}$Al \index{isotopes!26Al}\index{isotopes!53Mn} and $^{53}$Mn; the fact that these radioactive elements were still not decayed when those solids formed sets important constraints to the time interval between the latest nucleosynthesis event near the forming Sun and the actual condensation of solid bodies in the young solar system. This book will discuss these examples in detail, and illustrate the contributions of radioactivity studies to the subject of cosmic nucleosynthesis.

 \begin{figure} %%%%%%%%%%%%%%%%%%%%%%%%%%%%%%%%%%%
  \includegraphics[width=\textwidth]{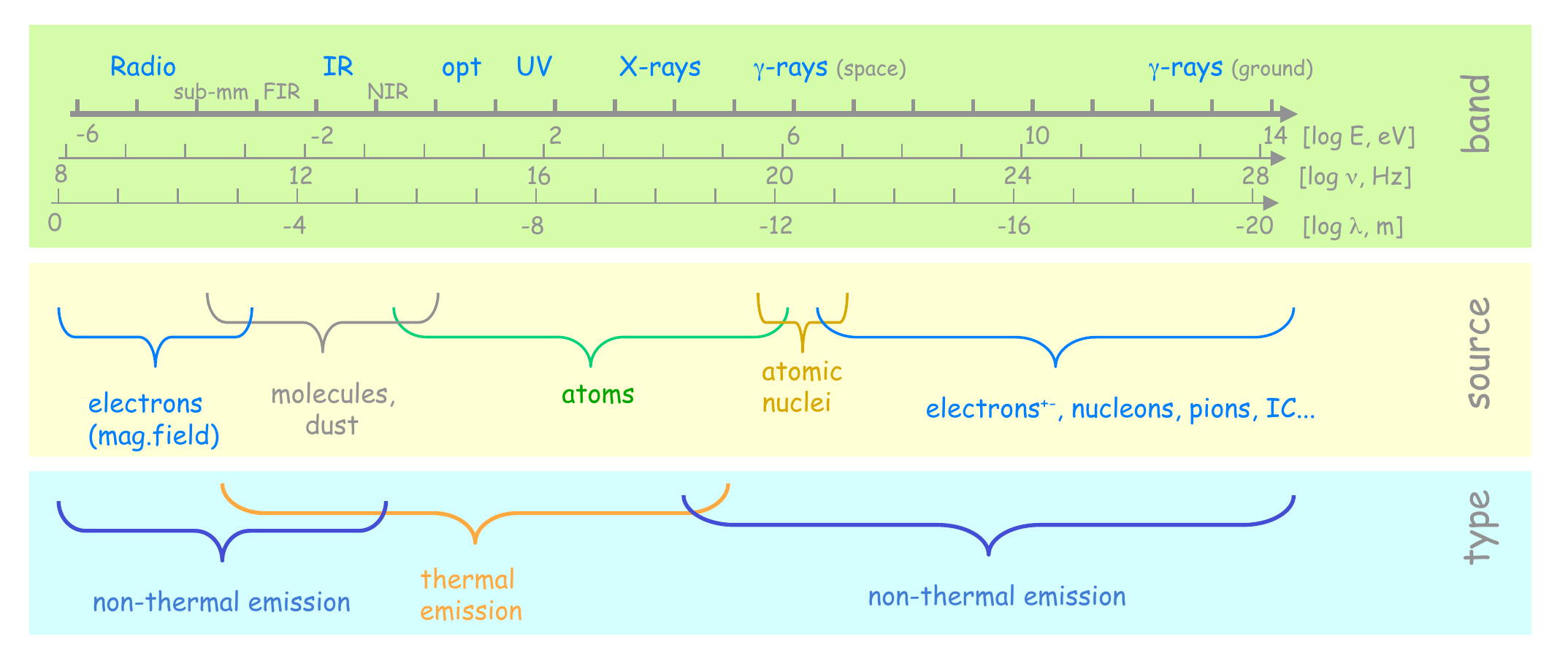}
  \caption{The electromagnetic spectrum \index{spectrum!electromagnetic}\index{electromagnetic radiation}  of candidate astronomical measurements ranges across more than twenty orders of magnitude. Not all are easily accessible. Information categories of thermal and non-thermal, and of molecular, atomic, nuclear, and elementary-particle physics origins of cosmic radiation extends over different parts of this broad spectrum. Nuclear physics is accessible in a small band (0.1-10 MeV) only, and is made difficult by penetrating radiation and by large instrumental backgrounds from cosmic-ray interactions.}
  \label{fig_1_emSpectrum}
\end{figure} %%%%%%%%%%%%%%%%%%%%%%%%%%%%%%%%%%%%%%

\section{Observations of Cosmic Radioactive Isotopes} %%%%%%%%%%%%%%%%%%%%%%%%%%%%%%%%
Astronomy has expanded beyond the narrow optical band into \emph{new astronomies} \index{astronomy} in the past decades. By now, we are familiar with telescopes measuring radio emission through infrared emission towards the long wavelength end, and ultraviolet, X-ray, and $\gamma$-ray emission towards the short wavelength end (see Fig.~\ref{fig_1_emSpectrum}). \index{spectrum!electromagnetic} The physical origins of radiation are different in different bands. Thermal radiation \index{radiation!thermal} dominates emission from cosmic objects in the middle region of the electromagnetic spectrum, from a few 10K cold molecular clouds at radio wavelengths through dust and stars up to hot interstellar gas radiating X-rays. \index{radiation!non-thermal} Non-thermal emission is characteristic for the wavelength extremes, both at radio and $\gamma$-ray energies. Characteristic spectral lines originate from atomic shell \index{spectrum!atomic}  electrons over most of the spectrum; nuclear lines are visible only in roughly two decades of the spectrum \index{spectrum!nuclear}  at 0.1--10~MeV. Few exceptional lines arise at high energy from annihilations \index{spectrum!annihilation} of \index{positron}\index{pion} positrons and pions. Thus, cosmic \emph{elements} can be observed in a wide astronomical range, while \emph{isotopes} are observed almost exclusively through $\sim$MeV $\gamma$-rays (see Fig.~\ref{fig_1_emSpectrum} for exceptions). Note that nucleosynthesis reactions occur among isotopes, so that this is the information of interest when we wish to investigate cosmic nucleosynthesis environment properties.

%%%%%%%%%%%%%%%%%%%%%%%%%%%%%%%%%%%%%%%%%%%%%%
\begin{figure}
%\centerline{
  \includegraphics[width=0.6\textwidth]{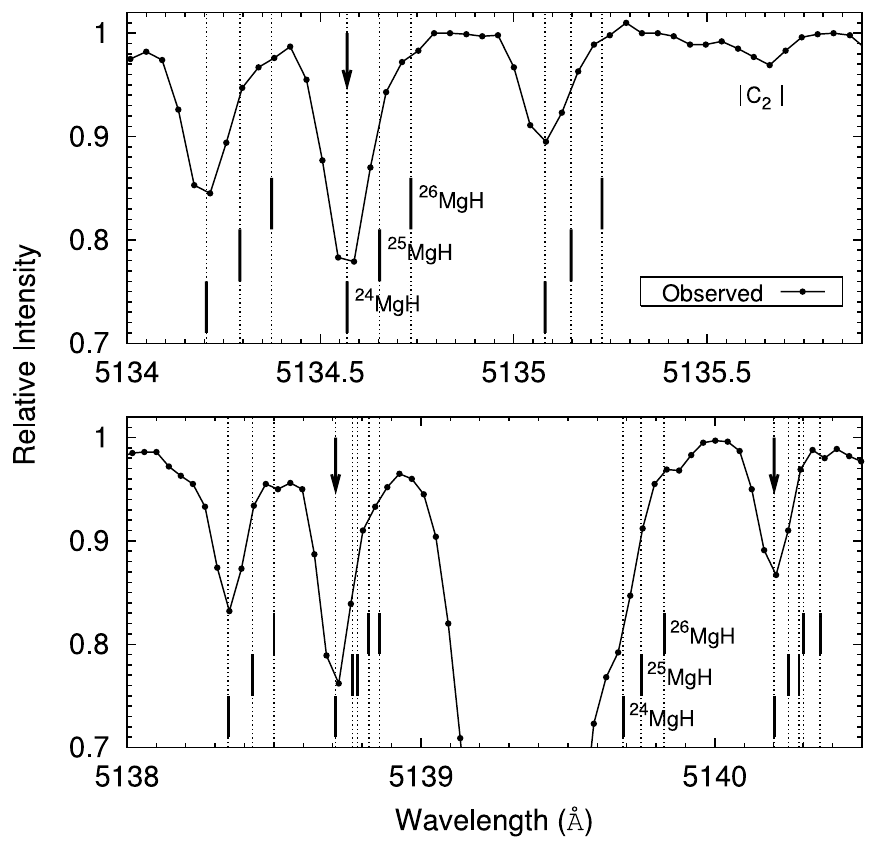}  %}
  \sidecaption[t]
  \caption{Example of an absorption-line spectrum \index{spectrum!absorption lines} of a cool star with a present-generation optical telescope, such as shown in Fig.~\ref{fig_1_ESO}. Molecular lines have isotopic shifts, which can be recognized as changes in line shapes, as resulting from the isotopic abundance ratio. \citep[from][]{2004ApJ...603..697Y}  }
  \label{fig_1_stellar_spectrosscopy}
\end{figure}
\begin{figure}
%\centerline{
  \includegraphics[width=0.6\textwidth]{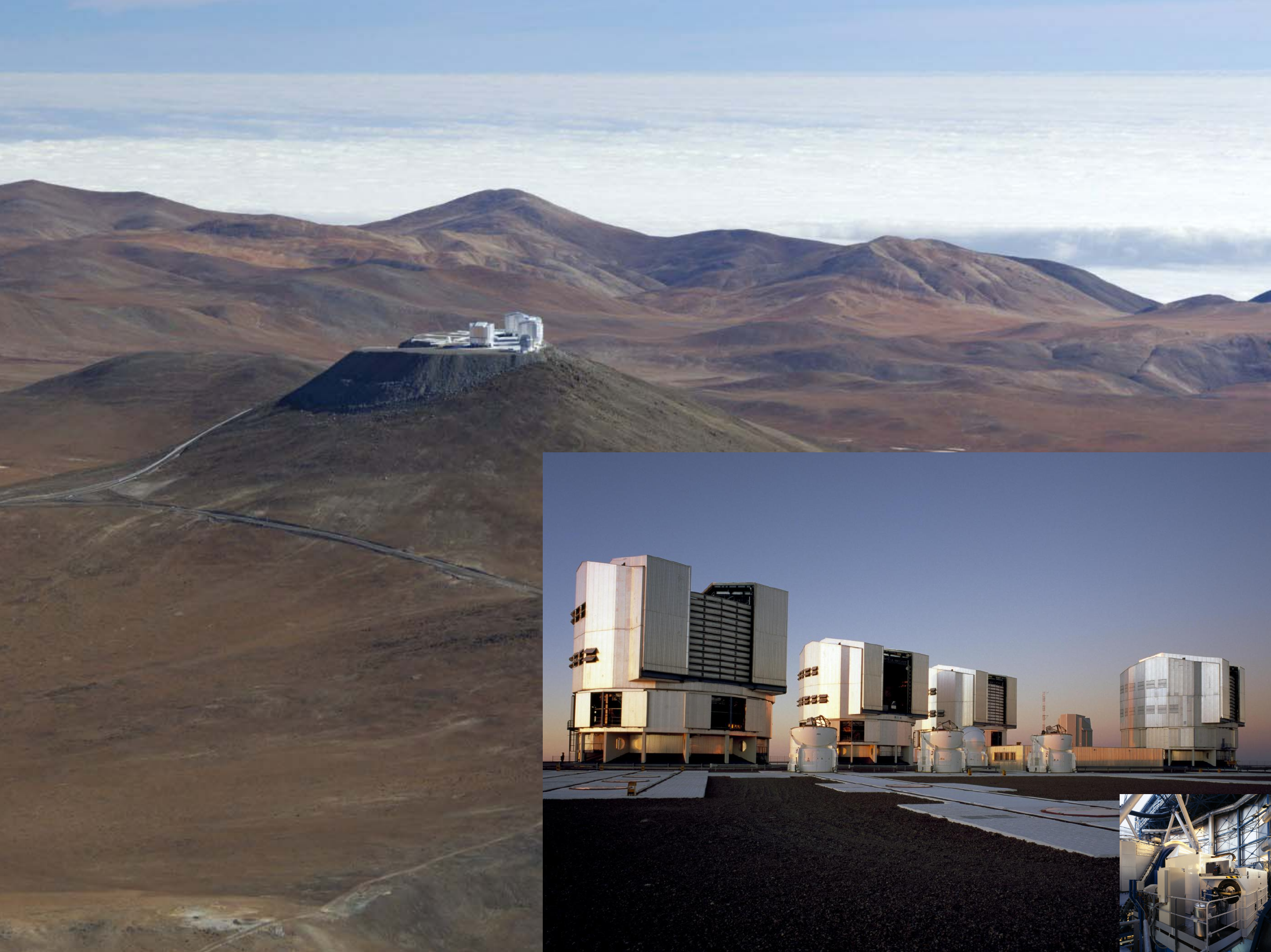}
 \sidecaption[t]
  \caption{The Very Large Telescope (VLT) on \index{Very Large Telescope (VLT)} Mount Paranal in Chile, with four telescopes (lower right), is one of the modern optical instruments. Equipped with high-resolution spectrographs such as FLAMES (insert lower right), absorption-line spectroscopy of stars in nearby galaxies can be made. (Figures ESO) }
  \label{fig_1_ESO}
\end{figure}

Only few elements such as technetium (Tc) \index{elements!Tc} \index{isotopes!98Tc} do not have any stable isotope; therefore, elemental photospheric absorption and emission line spectroscopy, the backbone of astronomical studies of cosmic nucleosynthesis, have very limited application in astronomy with radioactivities. This is about to change currently, as spectroscopic devices in the optical and lower-energy regime approach resolutions sufficient to resolve fine structure lines, thus enabling isotopic abundance studies. Observational studies of cosmic radioactivities are best performed by techniques which intrinsically obtain isotopic information. These are:
\begin{itemize}
\item 
precision mass spectroscopy \index{mass!spectroscopy} in terrestrial laboratories, which has been combined with sophisticated radiochemistry to extract meteoritic components originating from outside the solar system \index{meteorites}
\item 
spectroscopy of characteristic $\gamma$-ray lines \index{gamma-ray lines} \index{spectrum!gamma-ray} emitted upon radioactive decay in cosmic environments
\end{itemize}

\begin{figure}
  \centerline{
  \includegraphics[width=0.7\textwidth]{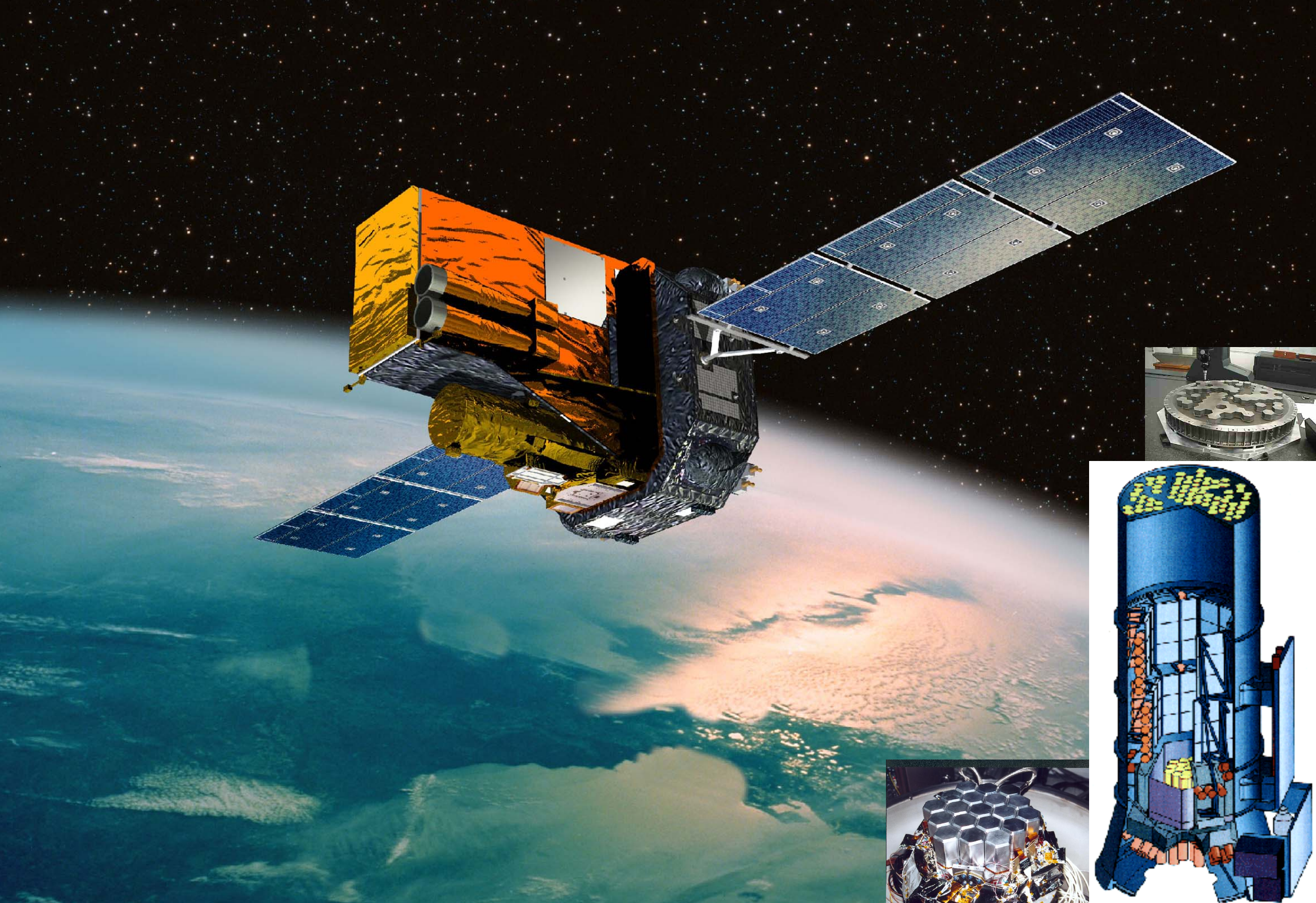}} % side-b-side: 0.49/0.45 width
% \sidecaption[t]
  \caption{Example of a present-generation space-borne $\gamma$-ray telescope. The INTEGRAL satellite \index{INTEGRAL} (artist view picture, ESA) has as one of its two main telescopes a spectrometer SPI, shown at the lower-right schematically with its 19-detector Ge camera and the tungsten mask for imaging by casting a shadow onto the camera.  Space-based instruments of this kind have been used to directly record characteristic $\gamma$-ray lines from the decay of unstable isotopes near sites of current-epoch cosmic element formation. }
 \label{fig_1_gamma}
\end{figure}

\begin{figure}
 \sidecaption[t]
 %\centerline{
  \includegraphics[width=0.6\textwidth]{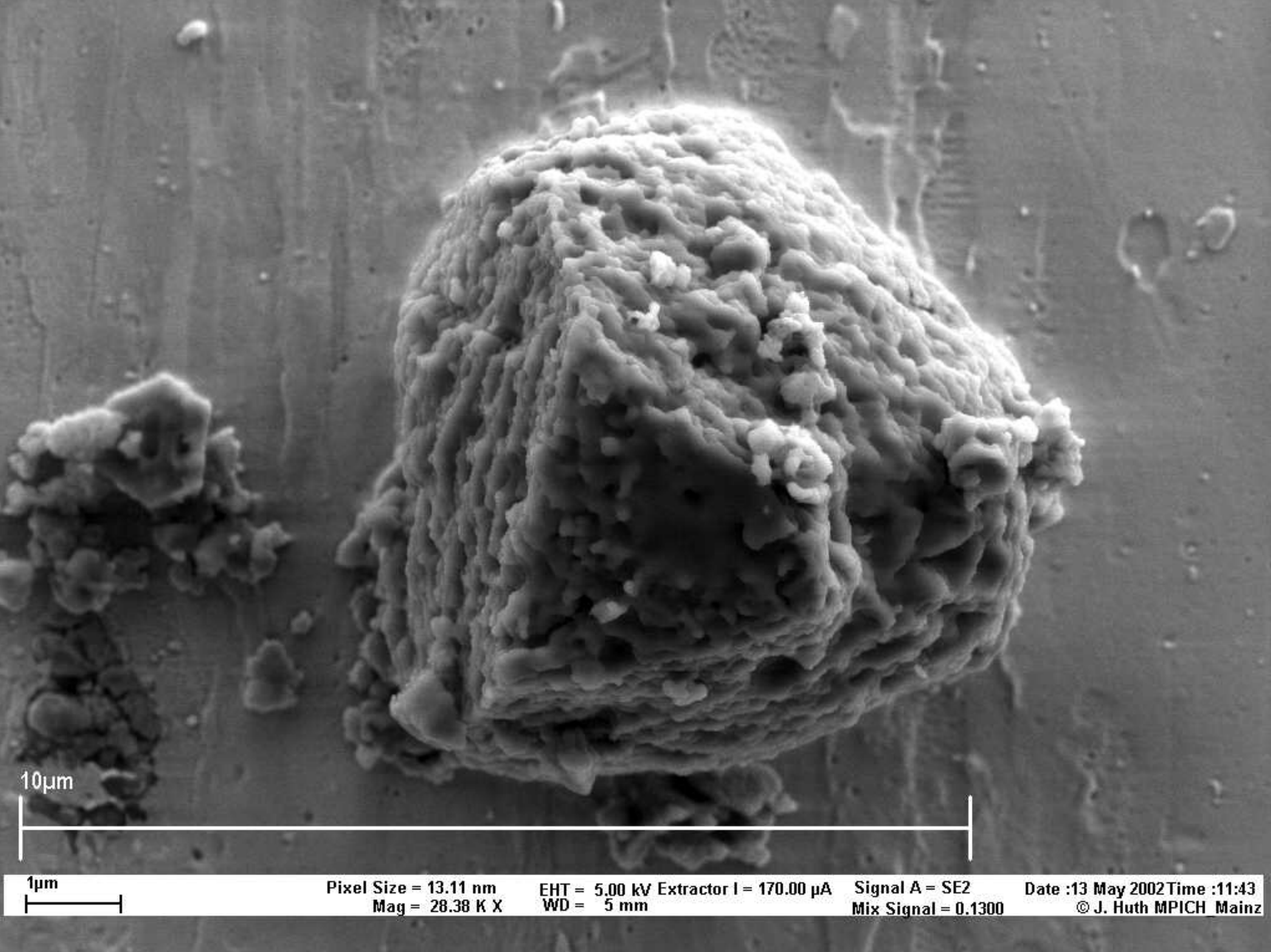}% 0.6 and 0.604 of same size}
  \caption{Meteoritic inclusions such as this SiC grain \index{meteorites} are recognized as dust formed near a cosmic nucleosynthesis source outside the solar system, from their large isotopic anomalies, which cannot be explained by interstellar nor solar-system processing but are reminiscent of cosmic nucleosynthesis sites. Having condensed in the envelope of a source of new isotopes, laboratory mass spectroscopy can reveal isotopic composition for many elements, thus providing a remote probe of one cosmic nucleosynthesis source.}
 \label{fig_1_grain}
\end{figure}

Both these {\it astronomical disciplines} have a relatively young history. They encounter some limitations due to their basic methods of how astronomical information is obtained:
\begin{itemize}
\item
Precision mass spectrometry of meteorites \index{meteorites} for astronomy with radioactivity began about 1960 with a new discovery of  now extinct radioactivity within the young solar system.  By heating a bulk meteorite sample the presence of \index{isotopes!129Xe} \index{isotopes!129I} excess  $^{129}$Xe was clearly demonstrated, and attributed to trapped gas enriched in $^{129}$I at the time of formation of this meteorite, which from mineralogical arguments is determined to be the early solar system epoch about 4.6~Gy ago \citep{PhysRevLett.4.8}. This was the first evidence that the matter from which the solar system formed contained radioactive nuclei whose half-lives are too short to be able to survive from that time until today ($^{129}$I decays to $^{129}$Xe within 1.7~10$^7$y).  Isotopic anomalies found in such \emph{extra-solar} inclusions, e.g. for C and O isotopes, range over four orders of magnitude for such \emph{star dust} grains as shown in Fig.~\ref{fig_1_grain} \index{stardust} \citep{1998AREPS..26..147Z}, while isotopic-composition variations among bulk meteoritic-material samples are a few percent at most. The measurements are characterized by an amazing sensitivity and precision, clearly resolving isotopes and counting single atoms at ppb levels to determine isotopic ratios of such rare species with high accuracy. 
This astronomy in terrestrial laboratories is now an established part of astrophysics \citep[see][for a recent review]{2004ARA&A..42...39C}. Studies are limited only by sample preparation and by the extraction techniques evaporizing dust grain surfaces for subsequent mass spectrometry. Substantial bias of the technique arises from the preparation of suitable samples, extracting meteoritic material with increased fractions of non-solar material and thus increasing the signal against background from solar-system isotopic composition. 
In general, this favors the hardest and least-resolvable meteoritic inclusions of most-refractory minerals, and noble gases included in microscopic cavities. Furthermore, the origin of non-solar dust, i.e. the journey from its formation in stellar envelopes or interstellar gas up to inclusion in meteorites which found their way to Earth, remains subject to theoretical modeling based on the observed grain composition and morphology plus (uncertain) theories of cosmic dust formation \citep{1998AREPS..26..147Z}. 
\item
Characteristic $\gamma$-ray lines \index{gamma-ray lines} from cosmic sources were not known until the 1960ies, when spaceflight and its investigations of the near-earth space radiation environment had stimulated measurements of $\gamma$-rays. The discovery of a cosmic $\gamma$-ray line feature near 0.5~MeV from the direction towards the center of our Galaxy in 1972 \citep{1972ApJ...172L...1J} stimulated balloon and satellite experiments for cosmic $\gamma$-ray line spectroscopy. By now  and with current experiments such as the INTEGRAL mission of ESA shown in Fig.~\ref{fig_1_gamma}, this technique has established an \index{astronomy} astronomical discipline of its own, the window of electromagnetic radiation clearly attributed to specific isotopes  \citep{2006NuPhA.777...70D}. 
Decay of the isotopes $^{26}$Al, $^{60}$Fe, $^{44}$Ti, $^{57}$Ni, and $^{56}$Ni \index{isotopes!26Al} \index{isotopes!44Ti} \index{isotopes!60Fe} \index{isotopes!57Ni} \index{isotopes!56Ni} in distant cosmic sites is an established fact, and astrophysical studies make use of such measurements. The downsides of those experiments is the rather poor resolution by astronomy standards (on the order of degrees), and the sensitivity limitations due to large instrumental backgrounds, which effectively only shows the few brightest sources of cosmic $\gamma$-rays until now \citep[see][for a discussion of achievements and limitations]{2006NuPhA.777...70D}. %(ref. review Diehl, von Ballmoos, Prantzos NuclPhys777, 2007).
\end{itemize}

Despite their youth and limitations, both methods to address cosmic radioactivities share a rather direct access to isotopic information, unlike other fields of astronomy. From a combination of all available astronomical methods, the study of cosmic nucleosynthesis will continue to advance towards a truly astrophysical decomposition of the processes and their interplays. This book describes where and how specific astronomical messages from cosmic radioactivity help to complement these studies.

\section{The Structure of this Book}
This book begins with a basic introduction in Chapter~2, written by D.D. Clayton, one of the founders of the field, who also contributed a fundamental textbook on stellar astrophysics and many stimulating ideas that determined the course of this field, in particular the need for a concerted observational effort to understand nucleosynthesis from both measurements of electromagnetic radiation and meteoritic material. This Introduction and Overview is followed by discussions of the specific sources of cosmic radioactivity in Chapters 3-5, i.e. stars in their hydrostatic interiors, massive stars and their core-collapse supernova explosions, and binary-star interactions including thermonuclear supernova explosions. These Chapters describe our current theoretical models and understanding of each of the cosmic sources of radioactivities, and guide these through references to key observations. Then we turn our perspective directly towards the observational side, and present prominent locations of radioactivities as observed (Chapter~6 addresses the Solar System, Chapter~7 more broadly then the different diffuse radioactivities seen in interstellar gas), and discuss how the data on cosmic radioactivities shape our astrophysical modeling of cosmic nucleosynthesis. The book is completed by a survey of a few tools which are characteristic of the field (in Chapters 8-10), and Appendices holding useful tables, a timeline of milestones of the field, and a glossary of key terms of Astronomy with Radioactivities to guide readers through the basic terminology. 
Beyond the general guidance of Chapter~2, subsequent chapters can be read in any sequence suitable to the interests and curiosities of the reader, as we provide cross references throughout the book. Enjoy the ride!

%%%%%%%%%%%%%%%%%%%%%% end of Chapter %%%%%%%%%%%%%%%%%%%%%%
% now the collected references for this chapter
%%%%\input{ch1/references_1}

\bibliographystyle{spbasic}
%\bibliography{rod-references}  % for first processing with BibTex

% end of chapter

\section*{Acknowledgements}

Many colleagues have contributed to this field, yet are not explicitly co-authors of this book. The choice of book authors grew out of opportunities around 2008, being aware that there were many more who could (and probably would) have joined into this adventure. All book authors are grateful for stimulating discussions and for collaborative work and publications shared with many dear friends and colleagues. We explicitly would like to mention here 
Sachiko	Amari,
Peter von	Ballmoos,
John	Beacom,
Peter	Biermann,
Bob	Binns,
Andreas Burkert, 
Roger 	Chevalier,
John Cowan, 
John	Danziger,
Mounib El Eid, 
Bruce 	Elmegreen,
Brian	Fields,
Claes	Fransson,
Roberto	Gallino,
Neil Gehrels, 
Matthieu 	Gounelle,
Alexander	 Heger,
Wolfgang	Hillebrandt,
Rob Hoffman, 
Peter	Hoppe,
Christian Illiadis, 
Anatoli	Iyudin,
Thomas Janka, 
Franz K\"appeler,
J\"urgen	Kn\"odlseder,
Gunther	Korschinek,
Karl-Ludwig	Kratz,
Pavel	Kroupa,
Karlheinz Langanke, 
John 	Lattanzio,
Bruno Leibundgut, 
Mark	 Leising,
Marco	Limongi,
G\"unther W.	Lugmair,
Gabriel Martinez-Pinedo,
Bradley	Meyer,
Georges	Meynet,
Peter	Milne,
Yuko	 Mochizuki,
Thierry	Montmerle,
Nami	Mowlavi,
Ewald M\"uller,
Ken'ichi Nomoto,
Uli	Ott,
Igor Panov, 
Etienne Parizot, 
Volker	Sch\"onfelder,
David	Smith,
Andrew Strong, 
Vincent	Tatischeff,
Lih-Sin	The,
James Truran,
Jacco	Vink,
Stan	Woosley,
Hans	Zinnecker,
and 
Ernst	 Zinner.
Thank you all.

This book is the result of team work with co-authors 
		Dieter Hartmann, 	
Nikos	Prantzos,
		Donald D. Clayton, 	
Maria	Lugaro, 	
				Alessandro Chieffi, 
Friedel	Thielemann, 
Raphael	Hirschi, 
		Matthias Liebend\"orfer,					
		Jordi Isern,	
		Margarida Hernanz,	
Jordi	Jose,
				Maurizio Busso	,
Thomas	Rauscher,
		Michael Wiescher, 	
Gottfried	Kanbach,
	and	Larry Nittler.
				I am grateful for their patience and endurance in this project, and their valuable contributions to this book.
			
 % Intro
 % \include{ch2/chapter_2} % Role of Radioactivities 

 % \include{part_2} % Sources of Isotopes
%  \include{ch3/chapter_3} % stars
% \include{ch4/chapter_4} % massive stars and cc-Sne
%  \include{ch5/chapter_5} % binary systems

% \include{part_3} % Special Places to see isotopes
% \include{ch6/chapter_6} % solar system
% \include{ch7/chapter_7} % diffuse radioactivities

%\include{part_4} % tools
% \include{ch8/chapter_8} % tools - simulating stars and supernovae
% \include{ch9/chapter_9} % tools - - nuclear reactions
% \include{ch10/chapter_10} % tools - instruments
%\include{chapter_12} % tools - databases

%% At end of each Chapter, the following commands handle references:
%%
%%\bibliographystyle{spbasic}
%%\bibliography{ch_X_refs}  % for ch_X_refs.bib file

%\include{part_5} % Closure

\backmatter%%%%%%%%%%%%%%%%%%%%%%%%%%%%%%%%%%%%%%%%%%%%%%%%%%%%%%%
%\setcounter{chapter}{10} %
%\include{outlook} % perspectives and outlook
%\appendix
%\include{Appendices/App_ChemEv}
%\include{Appendices/radioisotopes}
%\include{Appendices/timeline}
%\include{Appendices/glossary}
%%%%%%%%%%%%%%%%%%%%%%%%%%%%%%%%%%%%%%%%%%%%%%%%%%%%%%%%%%%%%%%%%%%%%%

\end{document}